\documentclass[runningheads,a4paper,11pt]{llncs}
\usepackage[text={6.5in,9in},centering]{geometry}

\usepackage[T1]{fontenc}
\usepackage[utf8]{inputenc}

\usepackage{amssymb}
\usepackage[english]{babel}
\usepackage{graphicx}
\usepackage{xspace, url, verbatim, array, float}
\usepackage[mathscr]{eucal}
\usepackage{comment}
\usepackage[version=3]{mhchem}
\usepackage{color}
\usepackage{cite}

\graphicspath{{./graphics/}}

\usepackage{dcolumn}
\newcolumntype{d}[1]{D{.}{.}{#1}}
\newcolumntype{C}{>{$}c<{$}}  %

\newcommand\cD{\mathcal{D}} 
 \newcommand\cM{\mathcal{M}}
\newcommand\cN{\mathcal{N}} \newcommand\cT{\mathcal{T}}

\newcommand\abs[1]{\left\lvert#1\right\rvert}

\newcommand\Rset{\mathbb{R}}
\newcommand\WITH{\,:\,}
\newcommand\prior[1]{\ensuremath{P_{\text{#1}}}} %

\newcommand\etal{\emph{et~al.\@}\xspace}

\newcommand\Prob{\mathbb{P}}

\newcommand\MA{\ensuremath{\mathit{MA}}}

\newcommand\mulossmass{\ensuremath{\mu_{\text{ls}}}}
\newcommand\sigmalossmass{\ensuremath{\sigma_{\text{ls}}}}
\newcommand\sigmamass{\ensuremath{\sigma_{\text{m}}}}
\newcommand\minint{\ensuremath{x_{\text{i}}}}
\newcommand\alphaint{\ensuremath{\alpha_{\text{i}}}}
\DeclareMathOperator{\erf}{erf}

\begin{document}

\title{Fragmentation Trees Reloaded}

\author{Kai Dührkop and Sebastian B\"ocker}

\institute{Chair for Bioinformatics, Friedrich-Schiller-University, Jena,
  Germany, \url{{kai.duehrkop,sebastian.boecker}@uni-jena.de}}

\date{\today}

\maketitle

\begin{abstract}
Metabolites, small molecules that are involved in cellular reactions, provide
a direct functional signature of cellular state.  Untargeted metabolomics
experiments usually relies on tandem mass spectrometry to identify the
thousands of compounds in a biological sample.  Today, the vast majority of
metabolites remain unknown.  Fragmentation trees have become a powerful tool
for the interpretation of tandem mass spectrometry data of small molecules.
These trees are found by combinatorial optimization, and aim at explaining
the experimental data via fragmentation cascades.  To obtain biochemically
meaningful results requires an elaborate optimization function.

We present a new scoring for computing fragmentation trees, transforming the
combinatorial optimization into a maximum a~posteriori estimator.  We
demonstrate the superiority of the new scoring for two tasks: Both for the de
novo identification of molecular formulas of unknown compounds, and for
searching a database for structurally similar compounds, our methods performs
significantly better than the previous scoring, as well as other methods for
this task.  Our method can expedite the workflow for untargeted metabolomics,
allowing researchers to investigate unknowns using automated computational
methods.
\end{abstract}

\section{Introduction}

Metabolites, small molecules that are involved in cellular reactions, provide
a direct functional signature of cellular state complementary to the
information obtained from genes, transcripts and proteins.  Research in the
field of metabolomics can give insight for biomarkers detection, cellular
biochemistry, and disease pathogenesis~\cite{yanes10metabolic,
  baker11metabolomics, patti12metabolomics, thaker13identifying}; whereas
natural product research screens metabolites for novel drug
leads~\cite{cooper11fix, hufsky14new}.  With advances in mass spectrometry
instrumentation, it is now possible to detect thousands of metabolites
simultaneously from a biological sample.

Metabolomics experiments often use a targeted approach in which only a
specified list of metabolites is measured.  This setup allows profiling these
metabolites with high speed, minimal effort and limited resources over a
large number of samples.  Unfortunately, the vast majority of metabolites
remain unknown~\cite{baker11metabolomics, patti12metabolomics}, and this is
particularly the case for non-model organisms and secondary metabolites.  The
structural diversity of metabolites is extraordinarily large; in almost all
cases, we cannot deduce the structure of metabolites from genome sequences.
To this end, untargeted metabolomics comprehensively compares the intensities
of thousands of metabolite peaks between two or more samples.  Here, a major
challenge is to determine the identities of those peaks that exhibit some
fold change.  For this, tandem mass spectrometry data of the
compounds is usually searched against a spectral library such as
Massbank~\cite{horai10massbank} or the Human Metabolome
Database~\cite{wishart09hmdb}.

Only few computational methods exist that target compounds not contained in a
spectral library~\cite{scheubert13computational, hufsky14computational}: In
particular, certain methods try to replace spectral libraries by the more
comprehensive molecular structure databases for
searching~\cite{allen14competitive, gerlich13metfusion, heinonen12metabolite,
  wolf10in-silico}.  But these methods must fail for compounds not present in
a structure database.  Methods for predicting the molecular formula of an
unknown compound usually require data beyond tandem mass
spectra~\cite{menikarachchi12molfind, pluskal12highly,
  rojas-cherto11elemental, boecker09sirius}.  Fragmentation trees (FTs) were
introduced to fill this gap~\cite{boecker08towards}, and were later shown to
contain viable structural information about the unknown
compound~\cite{rasche11computing}.  FT computation does not require any
(spectral, structural, or other) databases.

In this paper, we report a systematic approach for choosing the fragmentation
tree that best explains the observed data, based on Bayesian analysis and a
maximum a~posteriori estimation.  As conjectured
in~\cite{rasche11computing} this results in a strong increase in FT quality,
as we evaluate using two derived measures: Both for the de novo
identification of molecular formulas of unknown compounds, and for searching
a database for chemically similar compounds, the new FTs perform
significantly better than state-of-the-art methods.

\begin{figure}[tb]
\centering
\includegraphics[width=0.98\linewidth]%
  {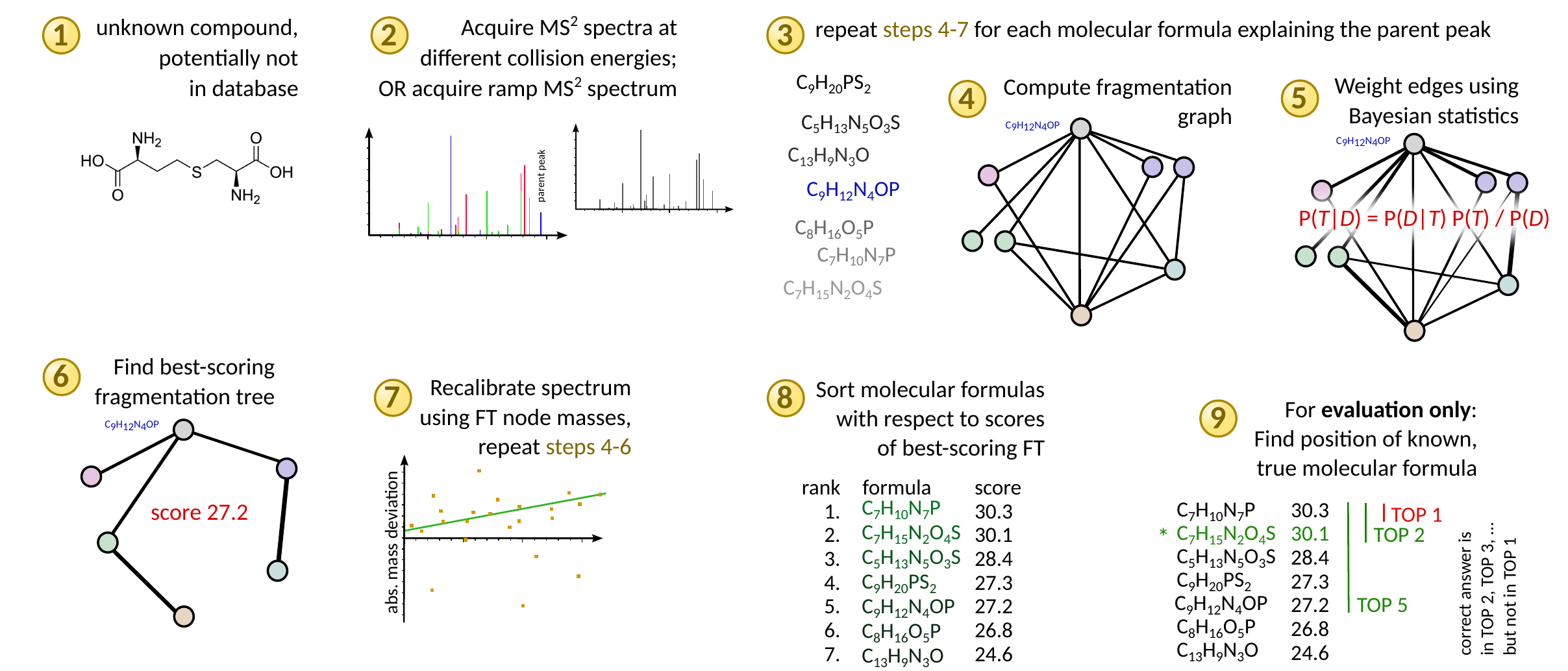}

\caption{Workflow of our method for computing FT to predict the correct molecular formula of the ion peak of MS/MS spectra.}
\label{fig:workflow}
\end{figure}

Our method will be made available on our website\footnote{\url{http://bio.informatik.uni-jena.de/software/}} 
as new version of the SIRIUS framework for MS and MS/MS analysis. See Fig.~\ref{fig:workflow} for a schematic workflow of our method.

\section{Fragmentation Trees}

First, we will formally introduce fragmentation trees, allowing us to
interpret fragmentation tree computation as a maximum a~posteriori estimation
in the next section.  Our \emph{data} $\cD = (\cM,I)$ is a measured
fragmentation spectrum with peak masses $\cM = \{m_1,\dots,m_L\}$ and peak
intensities $I: \cM \to \Rset_{> 0}$.  Masses are not measured with arbitrary
precision: To decide whether some theoretical molecular formula may coincide
with some measured peak, we use a relative mass accuracy parameter $\MA$
provided by the user.  Fragmentation spectra are relatively sparse: For any
interval of 1~Da in the spectrum, there is at most a few peaks present.  On
the other hand, we demand that the mass accuracy of the measurement is high,
say, 20~ppm or better.  To this end, almost all theoretical molecular
formula can explain \emph{at most one} peak in the measured spectrum.

A \emph{fragmentation tree} (FT) $\cT = (V,E)$ consists of a set of nodes $V$
which are molecular formulas over some alphabet of elements, and directed
edges (arcs) connecting these nodes.  All edges are directed away from the
root of the tree, and every node can be reached from the root via a unique
series of edges.  In small compound fragmentation, many fragments result from
fragmentation cascades, that is, series of subsequent fragmentation events;
these cascades are modeled by the tree structure of the FT.  Nodes of the FT
are molecular formulas of the unfragmented ion and its fragments; edges
correspond to losses.  For any FT, each molecular formula can appear at most
once as a node of the tree.  For an edge $(u,v) \in E$, $u-v$ is the
molecular formula of the corresponding loss; we demand that $u \ge v$ holds
(for each component) and, hence, $u-v \ge 0$.  Let $\mu(f)$ denote the
theoretical mass of the molecular formula~$f$ (either fragment or loss).
This will usually be the mass of the lightest naturally occurring isotope of
an element, such as $\mu(\ce{H}) = 1.007825$.

For a given FT, we can simulate a fragmentation spectrum (without
intensities), simply using the masses of all nodes' molecular formulas.  For
the inverse direction, a FT is supported by a fragmentation spectrum of a
(usually unknown) compound if, for every node of the tree, we find a peak in
the spectrum such that the mass difference between the molecular formula of
the node and the peak mass is below some user-defined threshold.  Not all
peaks of the fragmentation spectrum have to be explained by the tree, as we
also have to consider noise peaks in the spectrum.  We demand that every node
of the FT explains a \emph{unique} peak in the spectrum: no two nodes of the
tree may correspond to the same peak.  Allowing more than one node to explain
a peak, would violate the vast majority of observations: In theory, it is
possible that two fragments of a compound have different structure but very
similar mass, so that both fragments would explain the same peak.  In
practice, this situation is extremely rare and we can safely ignore it.

We now formalize our above considerations.  We say that a FT $\cT = (V,E)$ is
\emph{supported~by} the observed data $\cD = (\cM,I)$ if each node $v \in V$
is assigned a unique peak $m \in \cM$ in the fragmentation spectrum that is
within the chosen mass accuracy.  Furthermore, no two nodes are assigned the
same peak.  We denote the \emph{natural injective mapping} from the FT nodes
to the peaks by $m: V \to \cM$.  All peaks in the spectrum not assigned to a
node of the FT, are regarded as noise peaks.  Our task is to find a FT that
``best explains'' the observed data, where goodness-of-fit is measured by
some scoring function that matches FT and mass spectrum.

This formulation of the problem is not easily accessible by algorithmic
means; to this end, B\"ocker and Rasche~\cite{boecker08towards} gave an
alternative formulation which, for additive scorings, is equivalent to the
above: For each peak in the fragmentation spectrum, we find all molecular
formulas with mass difference sufficiently small.  These molecular formulas
are the nodes of a directed acyclic graph (DAG) called \emph{fragmentation
  graph}.  Nodes are colored so that all molecular formulas corresponding to
the same peak have the same color.  Recall that we must use at most one
vertex for each color (peak) in our FT.  Edges are inserted whenever one
molecular formula is a sub-formula of another.  Edges are appropriately
weighted using some score function.  It is straightforward to check that
there is a 1-1 correspondence between \emph{colorful} subtree, that use every
color in the graph at most once, and FTs supported by the data.  We search
for a colorful subtree of this graph that has maximum weight.  The underlying
computational problem has been coined \textsc{Maximum Colorful Subtree};
unfortunately, this problem is computationally hard~\cite{rauf13finding}.
Nevertheless, there exist a number of algorithms (both exact and heuristic)
to solve the problem in practice~\cite{rauf13finding,
  boecker08towards}.  In this paper, we will not cover any algorithmic
details of the problem; we solve our instances using Integer Linear
Programming (ILP) as described in~\cite{rauf13finding}.

\section{Maximum a~Posteriori Estimation}

Our maximum a~posteriori estimate roughly follows the scoring introduced by
Böcker and Rasche~\cite{boecker08towards}, further refined by Rasche
\etal~\cite{rasche11computing, rasche12identifying}.  These scorings were
motivated by stochastic considerations, but only in an informal way.  Here,
we will strictly model the problem as a maximum a~posteriori estimation, which
allows us to make sensible choices for the (hyper)parameters of the method.
Bayesian Statistics tell us that
\begin{equation} \label{equ:bayes}
  \Prob (\cT_j | \cD) = \frac{\Prob(\cD | \cT_j) \cdot
    \Prob(\cT_j)}{\Prob(\cD)} = \frac{\Prob(\cD | \cT_j) \cdot
    \Prob(\cT_j)}{\sum_i \Prob(\cD | \cT_i) \, \Prob(\cT_i)} \, ,
\end{equation}
where $\cD$ is the data (the measured spectrum) and $\cT_j$ are the models
(the candidate FTs).  We want to maximize the \emph{posterior probability}
$\Prob (\cT_j | \cD)$ which is equivalent to maximizing $\Prob(\cD | \cT)
\cdot \Prob(\cT)$ over all possible models~$\cT$.  Here, $\Prob(\cD | \cT)$
is the probability of the data given the model $\cT$, and $\Prob(\cT)$ is the
\emph{prior probability} of model~$\cT$, based on prior information that we
have about FTs without considering the actual data~$\cD$.  We have
considerable background information about the prior probability of any given
FT: For example, smaller losses are usually more frequent than larger losses,
and certain losses such as \ce{H2O} or \ce{CO} turn up very frequently.

We stressed repeatedly that we are interested in those FTs only that are
supported by the data.  To this end, we demand $\Prob(\cD|\cT) = 0$ and,
hence, $\Prob(\cT|\cD) = 0$ for any tree $\cT$ that is \emph{not supported by
  the data}~$\cD$.  In the following, we assume that each considered FT is
supported by the data.

We now introduce computations for prior probability and likelihood of the
tree.  Due to space constraints, we defer all details to the long version of this paper.

\subsection{Prior Probability of the Tree}

We first concentrate on the prior $\Prob(\cT)$.  We will not demand that
priors sum to one but only that the sum $\sum_i \Prob(\cT_i) \, \Prob(\cD |
\cT_i)$ converges, what is sufficient for optimizing $\Prob(\cT) \cdot
\Prob(\cD | \cT)$.  But this is obviously true: The number of models $\cT_i$
we are considering is finite, as we are only consider trees supported by the
data.  We assume that, for all trees of constant size, prior probabilities of
the nodes and edges of $\cT$ are independent so that
\[
  \Prob(\cT) = \Prob (\text{size $\abs{E}$ of the tree}) \cdot
  \prod\nolimits_{v \in V} \Prob(v) \cdot \prod\nolimits_{e \in E} \Prob(e) .
\]
Here, $\Prob(v)$ is the prior probability to see a particular \emph{fragment}
in a FT, and $\Prob(e)$ is the prior probability to see a particular
\emph{loss} in a FT.  The independence assumption is obviously violated in
reality, but allows us to come up with simple yet meaningful priors.  We can
simplify this equation, noting that every node of the tree except the root
has exactly one incoming edge.  For molecular formulas $u,v$ let
$\prior{edge}(u,v)$ be the prior that fragment $v$ \emph{and} loss $u-v$ are
simultaneously seen in the tree, and let $\prior{root}(u)$ be the prior that
the tree is rooted with molecular formula~$u$.  Then,
\begin{equation}
  \Prob(\cT) \varpropto \Prob (\text{size $\abs{E}$ of the tree}) \cdot
  \prior{root}(r) \cdot \prod_{(u,v) \in E} \prior{edge}(u,v)
\end{equation}
where $r$ is the root of~$\cT$.

\paragraph{Prior of the Root.}

We use the following uninformative prior to filter out structurally
impossible molecular formulas: For each compound, the sum of valences has to
be greater than or equal to twice the number of atoms minus
one~\cite{senior51partitions}.  This corresponds to a non-negative ring
double bond equivalent (RDBE) value.  In addition, we use five informative
priors: First, assume that the compound is not a radical, then the sum of
valences is even~\cite{senior51partitions}.  If the compound ion is
protonated, then the sum of valences of the ion is odd.  As both
intrinsically charged molecules and free radicals are comparatively rare, we
use prior $0.1$ for molecular formulas with even sum of valences, and $1$ for
all others.  Second, the ratio between hetero atoms and carbon atoms is
usually relatively small for biomolecules~\cite{kind07seven}.  We find that
this ratio becomes even more informative if we also exclude oxygen from the
hetero atoms.  We model the ``hetero minus oxygen to carbon ratio'' (HMOTCR)
using a uniform prior for small ratios, and a Pareto distribution for larger
ratios.  Third, for the ring double bond equivalent (RDBE), we observed that
the value $\text{RDBE}/{m^{2/3}}$ is roughly normal distributed, where $m$ is
the mass of the compound.  We use the density of the normal distribution as
the prior.  The last two priors penalize molecular formulas containing rare elements
as well as formulas containing phosphor atoms without oxygen or sulfur atoms (as $99 \%$ of the compounds in KEGG
that contain phosphor also contain oxygen or sulfur).
The root prior $\prior{root}(r)$ is the product of these five
priors.  We stress that informative priors never discard any molecular
formulas but rather, decrease the scores of these formulas. The root prior becomes less important as more peaks are contained in the spectrum (and nodes in the tree). But for compounds that do not fragment very well the root prior may help to identify the correct molecular formula.

\paragraph{Priors of Edges.}

The prior probability $\prior{edge}(u,v)$ of an edge $e = (u,v)$ is estimated
from different factors, namely prior knowledge about implausible (and
radical) losses, the mass of the loss, common losses, as well as common
fragments.  We first penalize \emph{implausible losses} of an edge $(u,v)$
using a prior $\prior{loss-impl}(u,v)$ on the loss $u-v$.  This is a small
list of losses that repeatedly turned up during our combinatorial
optimization, but that were rejected in the expert evaluation
in~\cite{rasche11computing}.  In particular, we penalize losses that contain
only nitrogen or only carbon; radical losses with certain exceptions; and few
losses from a list of losses generated by expert knowledge.  Since these are
losses that we \emph{do not want to see}, there appears to be no sensible way
to learn such implausible losses from the data.  Instead, we have to rely on
expert knowledge and evaluation of FTs computed by the method, to collect
this list.  Also, priors for such implausible losses were chosen
\emph{ad~hoc} as there appears to be no sensible way of learning such
penalties from the data.

Regarding the mass of a loss, we assume that large losses are less likely
than small losses.  Unfortunately, there is only a very small number of
annotated FTs available in the literature, and these are usually measured on
different instruments (and instrument types) using different experimental
setup and, hence, are mostly incomparable.  To this end, we chose to estimate
the loss mass distribution using FTs determined by our method.  Different
from~\cite{boecker08towards, rasche11computing, rasche12identifying} we do
not penalize the relative size of the mass but rather the mass itself, as
this allows for a more stringent incorporation of common losses.
Combinatorics dictates that there exists only a small number of losses below,
say, $30$~Da.  Besides certain common losses, this implies that the number of
small losses is also small, but increases rapidly until some maximum is
reached.  Beyond this mass, we find that the probability to observe a loss
drops rapidly in the beginning, but stays significantly above zero even for
large masses.  To model these observations, we use a log-normal distribution
as a classical example of a long-tailed distribution.

Some losses turn up more often than we would expect from the loss mass
distribution.  In~\cite{boecker08towards} a expert-curated list of common
losses was introduced, and this list was further refined
in~\cite{rasche11computing, rasche12identifying}.  Such hand-curated lists
can be incomplete and, worse, prior probabilities have to be chosen
\emph{ad~hoc}.  We chose to learn common losses and their prior
probabilities from our training data.

Similar to the root, we want to penalize molecular formulas with extreme
``hetero minus oxygen to carbon ratio'' (HMOTCR) and RDBE value of a
fragment.  As proposed in~\cite{boecker08towards} we do not penalize a child
if we have already penalized the parent, as both HMOTCR and RDBE values are
hereditary.  We set the prior to be the minimum value of one and
the ratio of the priors of child and parent.

For a FT to be informative, it is useful that the FT includes fragments of
small masses, even if the corresponding peaks have small intensities and,
possibly as a result, larger mass deviations. The molecular formula identification
of peaks with small masses is easier due to fewer possible 
explanations. Therefore, we add a prior that rewards fragments with small masses.

Finally, we noticed that certain fragments turn up repeatedly in FTs.  The
explanation for this observation is simple and is known to MS experts for
decades: Certain groups such as \ce{C6H4} (benzyne) or \ce{C4H7N} (pyrroline)
can be cleaved off as ions, leading to characteristic peaks in the mass
spectra.  But giving priors for both common losses \emph{and} common
fragments, clearly violates the independence assumption: If we know the
molecular formulas of a fragment and one of its losses, then this also tells
us the molecular formula of the child fragment.  To this end, we chose a
``cautious'' prior that rewards only few and small common fragments which
have been observed very often, whereas the vast majority of fragments receive
a flat prior.

\paragraph{Prior of the Tree Size.}

The FT we will compute should explain a large number of peaks; We want to favor large trees over small ones.  The priors we have introduced
so far do exactly the opposite: Many edges result in many probabilities we
have to multiply, and small trees are favored over large trees.  To this end,
we introduce one last prior: We assume $\Prob (\text{size $\abs{E}$ of the
  tree}) \varpropto \prior{tree-size}^{\abs{E}}$ where $\prior{tree-size} :=
\prior{tree-norm} \cdot \prior{tree-bonus}$.  Here, $\prior{tree-norm}$ is
chosen to counter the effects of the other priors on average, whereas
$\prior{tree-bonus}$ is $-0.5$ by default but can be increased by the user to favor larger trees.

\subsection{Likelihood of the Tree} \label{sec:likelihood}

Recall that each considered FT $\cT = (V,E)$ is supported by the data~$\cD =
(\cM,I)$.  This implies the existence of a natural injective mapping $m: V
\to \cM$: Each node $v \in V$ is assigned a unique peak $m(v)$ in the
fragmentation spectrum.  All peaks in the spectrum not assigned to a node of
the FT, are noise peaks and also contribute to the likelihood of the tree.

To simplify our computations, we assume independence between the measured
peaks in $\cM = \{m_1,\dots,m_L\}$, so $\Prob(\cD | \cT) = \prod_l \Prob(m_l
| \cT)$.  Here and in the following, $m_l$ refers both to the $l$-th peak and
to its mass.  Furthermore, we may assume that for each peak, the probability
of the tree to generate some peak depends only on the corresponding
hypothetical fragment, so $\Prob(m(v) | \cT) = \Prob(m(v) | v)$ for all $v
\in V$.  Then,
\[
  \Prob(\cD | \cT) = \prod_l \Prob(m_l | \cT) = \prod\nolimits_{v \in V}
  \Prob(m(v) | v) \cdot \Prob(\text{unassigned peaks} | \cT)
\]
for appropriately chosen $\Prob(m(v) | v)$.  Here, $\Prob(\text{unassigned
  peaks} | \cT)$ is the probability that all unassigned peaks $\cM - \{m(v)
\WITH v \in V\}$, which cannot be explained by $\cT$, are noise peaks.  We
assume that different noise peaks are again independent.

Unassigned peaks cannot be scored in the FT optimization, as only those nodes
and edges are scored that are actually part of the tree.  Note again that
each node is assigned a unique peak, and that no two nodes are assigned the
same peak.  We reach
\[
  \Prob(\cD | \cT) = \Prob(\text{all peaks are noise}) \cdot
  \prod_{v \in V} \frac{\Prob(m(v) | v)}{\Prob(\text{$m(v)$ is
      noise})}
\]
for appropriate $\Prob(m(v) | v)$.  Again, for fixed data~$\cD$, the
probability of all peaks being noise simultaneously is a constant, and can be
ignored in the optimization of $\Prob(\cT | \cD)$.

We will now show how to compute the probability of signal peaks and noise
peaks.  Currently, there exists no general model for the intensity of signal
peak in small compound MS.  Here, the problem is even harder, as we do not
know the fragment's molecular \emph{structure} but only its molecular
formula.  Similarly, there exists no sensible model for the mass of noise
peaks.  To this end, we will use only the peak mass to assess the probability
of signal peaks; and only peak intensity to assess the probability of noise
peaks.  The intensity of peak $m$ is $I(m)$; for brevity we write $I(v) :=
I(m(v))$.

\paragraph{Probability of Signal Peaks.}

It has been frequently observed that relative mass deviations are roughly
normally-distributed~\cite{jaitly06robust, zubarev07proper}.  We found this
to be the case for our datasets, too.  We assume that the instrument is
decently calibrated, then relative mass errors are distributed according to
$\cN(0,\sigmamass)$.  
We ignore that no mass errors above some threshold can be observed (truncated
normal distribution) as this has a negligible effect on our computations.
The probability to observe a peak with mass $m(v)$ for node/fragment $v$ can
be estimated as
\begin{equation}
  \Prob(m(v) | v) = \Prob \left( \abs{\cN(0,\sigmamass)} \ge \tfrac{\abs{m(v)
      - \mu(v)}}{\mu(v)} \right) = \erf \left( \tfrac{\abs{m(v) -
      \mu(v)}}{\sigmamass \sqrt{2} \mu(v)} \right).
\end{equation}
This is the two-sided probability that a mass deviation larger than the
observed relative mass deviation of peak $m(v)$ will occur by chance.  Here,
``$\erf$'' denotes the error function.

\paragraph{Probability of Noise Peaks.}

We can estimate the probability that a certain peak is noise, by observing
that noise with high intensity are much rarer than noise peaks with small
intensity.  Previous versions of FT calculation~\cite{boecker08towards,
  rasche11computing} implicitly assumed that noise peak intensities are
exponentially distributed.  For our data, we observe that with increasing
intensity, the probability to observe a noise peak of this intensity drops
rapidly in the beginning, but stays significantly above zero even for large
intensities.  This is an example of a long-tailed distribution, and we use
the Pareto distribution as a classical example of a long-tailed distribution.

Let $\minint$ be the peak intensity threshold used for peak picking.  The
probability density function of the Pareto distribution is $\alphaint
\minint^{\alphaint} / x^{\alphaint+1}$ for mass~$x$. $\alphaint$ is the shape 
parameter of the distribution and can be learned from data using a maximum likelihood estimator.
The probability of observing a noise peak $m$ with intensity $I$ or higher, is $\Prob(\text{$m$
  is noise}) = \alphaint \minint^{\alphaint} / I^{\alphaint+1}$.

\subsection{Posterior Probability of the Tree}

From the above we infer that
\begin{equation}
  \Prob(T) \cdot \Prob(\cT | \cD) \varpropto \prior{root}(r) \cdot \prod_{e
    \in E} (\prior{edge}(e) \cdot \prior{tree-size}) \cdot \prod_{v \in V}
  \left( \erf \left( \tfrac{\abs{m(v) - \mu(v)}}{\sigmamass \sqrt{2} \mu(v)}
  \right) \Bigm/ \tfrac{\alphaint
    \minint^{\alphaint}}{I(v)^{\alphaint+1}} \right)
\end{equation}
for FT $\cT = (V,E)$ with root $r \in V$.  This allows us to weight the edges
of the fragmentation graph: For each edge $(u,v)$ we set its edge weight
\begin{equation}
  w(u,v) := \log \prior{edge}(u,v) + \log \prior{tree-size} + \log \erf
  \left( \tfrac{\abs{m(v) - \mu(v)}}{\sigmamass \sqrt{2} \mu(v)} \right) -
  \log \tfrac{\alphaint \minint^{\alphaint}}{I(v)^{\alphaint+1}} .
\end{equation}
With these edge weights, the colorful subtree
of maximum weight corresponds to the FT with maximum posterior probability;
more precisely, ordering colorful subtrees with respect to their weight, is
equivalent to ordering the corresponding FTs by posterior probability.

\subsection{Hypothesis-driven Recalibration.}

To improve the quality of FTs, we have implemented a hypothesis-driven
recalibration~\cite{boecker08combinatorial}.  We are given one fragmentation
spectrum at a time.  For each candidate molecular formula explaining the
root, we compute a FT, and then use the theoretical masses of all nodes in
the FT as references to recalibrate the sample spectrum. We then compute the optimal FT for the recalibrated sample spectrum and the
candidate molecular formula, and use this score to evaluate which root
molecular formula best explains the data.  Then, the recalibration is
\emph{discarded}, returning to the original measured sample spectrum, and the
next root molecular formula is processed.

We note that our hypothesis-driven recalibration (HDR) is fundamentally
different from, say, the recalibration proposed
in~\cite{stravs13automatic}: Using HDR, each spectrum is recalibrated
individually, using each peaks best theoretical explanation as anchors for
the mass correction.  In this way, we do not require a homogeneous dataset of
mass spectra to start the recalibration process.

\section{Results}

\paragraph{Datasets.}

The GNPS dataset was downloaded from the GNPS database in December 2014
(\url{http://gnps.ucsd.edu}).  We analyze a total of 2\,006 non-peptide
compounds with mass below 1010 Da where mass spectra were recorded in positive
mode, and mass accuracy of the parent mass was below 10~ppm.  For each compound,
a single fragmentation spectra was recorded on an Agilent QTOF with
electrospray ionization.  The \emph{Agilent} dataset is available under the name ``MassHunter
Forensics/Toxicology PCDL'' (version B.04.01) from Agilent Technologies
(Santa Clara, CA, USA).  The commercial library has been cleaned by
idealizing peak masses and removing noise peaks, but Agilent provided us with
an uncorrected version of this dataset, which is used here.  For this
dataset, 2\,120 compounds fulfill the above criteria.  Fragmentation spectra
at collision energies 10, 20, and 40 eV were recorded on an Agilent 6500
Series QTOF system with electrospray ionization.  Only relative intensities
were recorded, so preprocessing was applied to merge spectra recorded at
different collision energies.
The masses of the compounds in both dataset range from 85 Da to 980 Da with an average mass of 340 Da.

Each dataset was split into two disjoint batches: The \textbf{\ce{CHNOPS}}
batch contains compounds that use solely elements \ce{CHNOPS} (GNPS: 1\,589
compounds, Agilent: 1\,540 compounds), whereas compounds from the
\textbf{contains \ce{FClBrI}} batch contain at least one atom from
\ce{FClBrI} (GNPS: 417, Agilent: 580).

\paragraph{Estimating the (Hyper)parameters.}

To apply our model to real data, we have to fit the (hyper)parameters for
priors and the likelihood estimation. We optimize hyperparameters in an iterative procedure, using FTs
from the previous round to determine parameters of the current.  See the
long version of this paper for all details.

\begin{figure}[tb]
\centering
\includegraphics[width=0.98\linewidth]{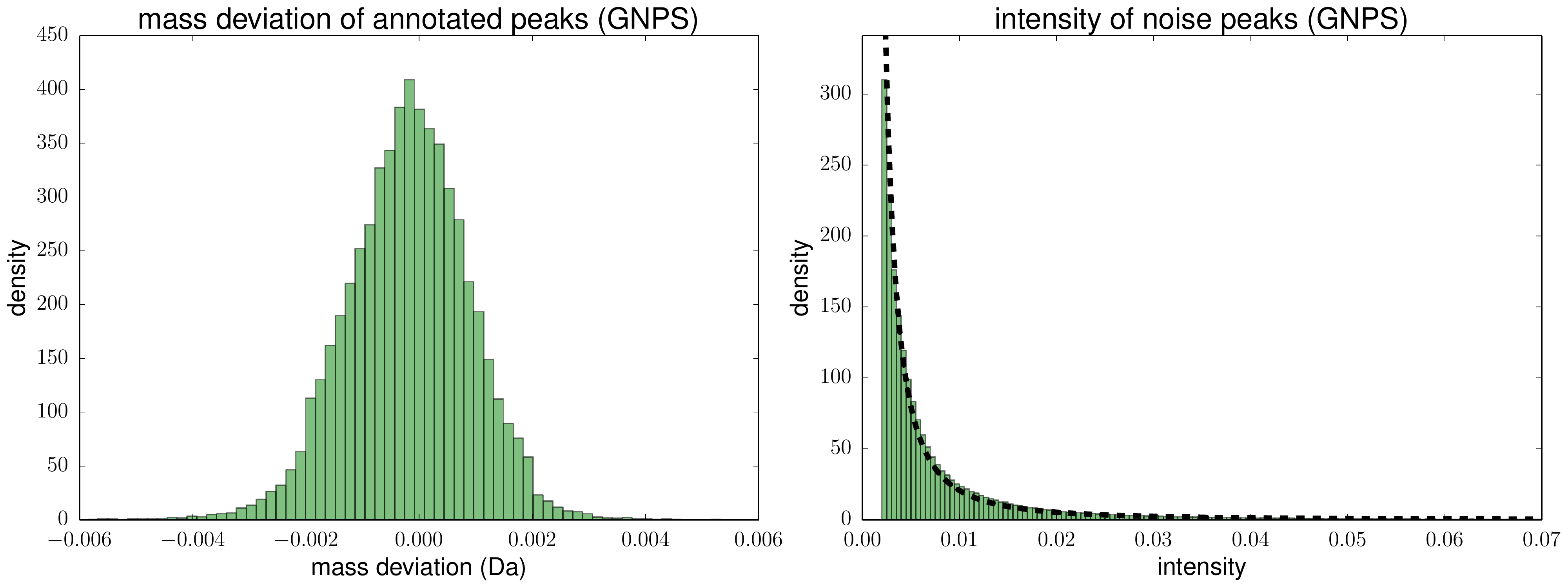}

\caption{Left: Normalized histogram of the mass error distribution.  Right: Normalized histogram of the noise peak intensity
  distribution and fitted Pareto distribution (dashed line).
  GNPS dataset.}
\label{fig:noise-and-accuracy}
\end{figure}

For both datasets, we observe that mass errors follow a normal distribution,
see Fig.~\ref{fig:noise-and-accuracy}.  By manual inspection, we estimate
$\MA = 10$~ppm.  The maximum likelihood estimation for our datasets leads to a normal distribution with $\sigmamass = 5.5$. 
But we find that using a higher standard deviation $\sigmamass = 10$ gives us better results due to the lower weight our scoring is giving to the mass deviation.
In both datasets, we observe an exponential decay of
noise peaks (i.e. peaks without an explanation for the parent molecular formula) with increasing intensity, see Fig.~\ref{fig:noise-and-accuracy}.
We estimate $\minint = 0.002$ and $\alphaint = 0.34$ for GNPS and $\minint = 0.005$, and $\alphaint = 0.5$ for Agilent.

\begin{figure}[tb]
\centering
\includegraphics[width=0.98\linewidth]%
  {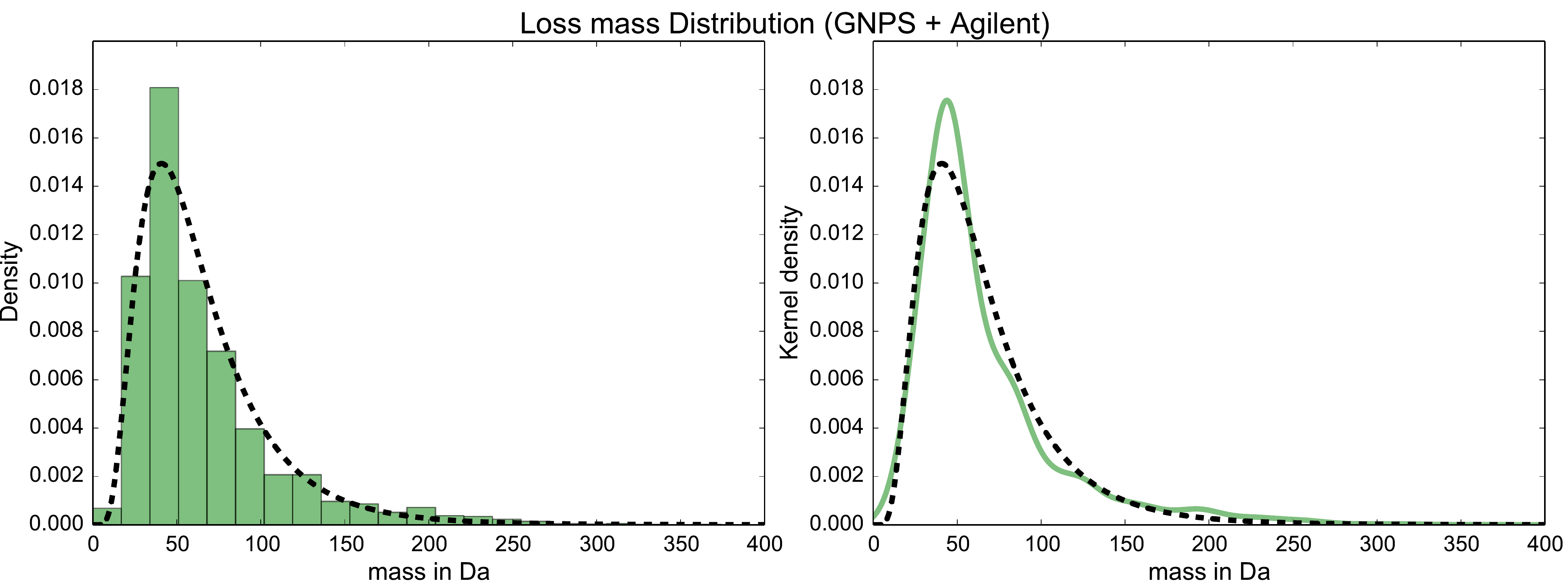}

\caption{Loss mass distribution, after the final round of parameter
  estimation.  Frequencies of the losses are weighted by the intensity of their peaks. The frequency of the identified common losses have
  been decreased to the value of the log-normal distribution.  Left:
  Normalized histogram for bin width $17$~Da. Right: Kernel density estimation. Black (dashed): Maximum likelihood estimate of the log-normal
  distribution.}
\label{fig:lossmass}
\end{figure}

Common losses are outliers, in the sense that their frequency is far higher
than we would expect for a loss of this mass.  During our iterative procedure we find 34 common losses, 13 of them were already listed in~\cite{boecker08towards,rasche11computing,rasche12identifying}, further 16 losses could be assigned to known structures. See Fig.~\ref{fig:lossmass} for
the agreement between the observed distribution of loss masses (corrected for
common losses) and the fitted log-normal distribution.  We estimate
$\mulossmass = 4.02$ and $\sigmalossmass = 0.31$ for the loss mass
distribution, with mode $e^{\mulossmass} = 55.84$ Da.

\paragraph{Evaluation Results.}

There is practically no way to determine the \emph{ground truth} of the
fragmentation process; even the comparison with fragmentation cascades
obtained using MS$^n$ data is not a satisfactory solution.  Manual evaluation
is very work-intensive and, hence, infeasible for the two large-scale
datasets considered here.  To this end, we evaluate the performance of our
method in answering a question where the true answer is known.

To identify the \emph{molecular formula} of a compound, we rank the FTs and,
hence, the molecular formulas according to the reached posterior
probabilities.  Besides mass accuracy and noise peak intensity, the user has
to provide the alphabet of elements the unknown compound can be made from.
For batch \ce{CHNOPS} we use this alphabet of elements without further
restrictions.  For batch ``contains \ce{FClBrI}'' we assume that we know
upfront which of the elements, besides \ce{CHNOPS}, \emph{may} be contained
in the compound.  Such information can be obtained from the isotope pattern
and the tandem mass spectrum using, say, machine learning (manuscript in
preparation).

\begin{figure}[tb]
\centering
\includegraphics[width=0.96\linewidth]{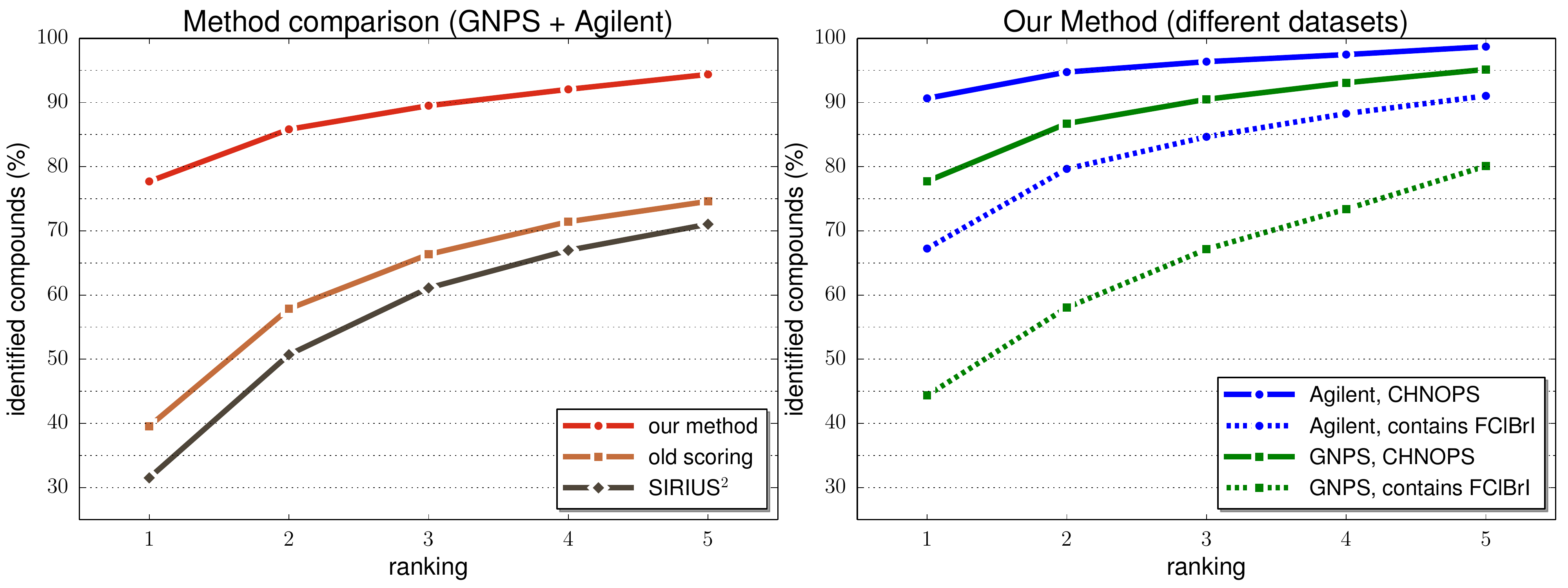}

\caption{Performance evaluation, percentage of instances (y-axis) where the
  correct molecular formula is present in the TOP~$k$ for $k=1,\dots,5$
  (x-axis).  Left: Performance evaluation for different
  methods on both datasets.  Methods are ``our method'' (the method
  presented here), ``old scoring'' (scores from \cite{rasche11computing,
    rasche12identifying} with ILP), ``SIRIUS$^2$'' (scores from
  \cite{rasche11computing, rasche12identifying} with DP).  
  Right: Performance for the two compound batches (\ce{CHNOPS} as
  solid line, ``contains \ce{FClBrI}'' as dashed line) and the two datasets
  (GNPS green, Agilent blue).}
\label{fig:performances}
\end{figure}

See Fig.~\ref{fig:performances} for the molecular formula prediction
performance of the method.  As expected, prediction is much harder for the
batch containing halogens.  Also, the new scoring significantly increases the
number of instances where we can recover the correct molecular formula.  We
evaluate our method both against the method from \cite{rasche11computing,
  rasche12identifying} as published there, using a dynamic programming (DP)
approach for finding the best FT; plus, scores from \cite{rasche11computing,
  rasche12identifying} together with the ILP from~\cite{rauf13finding}.

\begin{figure}[tb]
\centering
\includegraphics[width=0.96\linewidth]{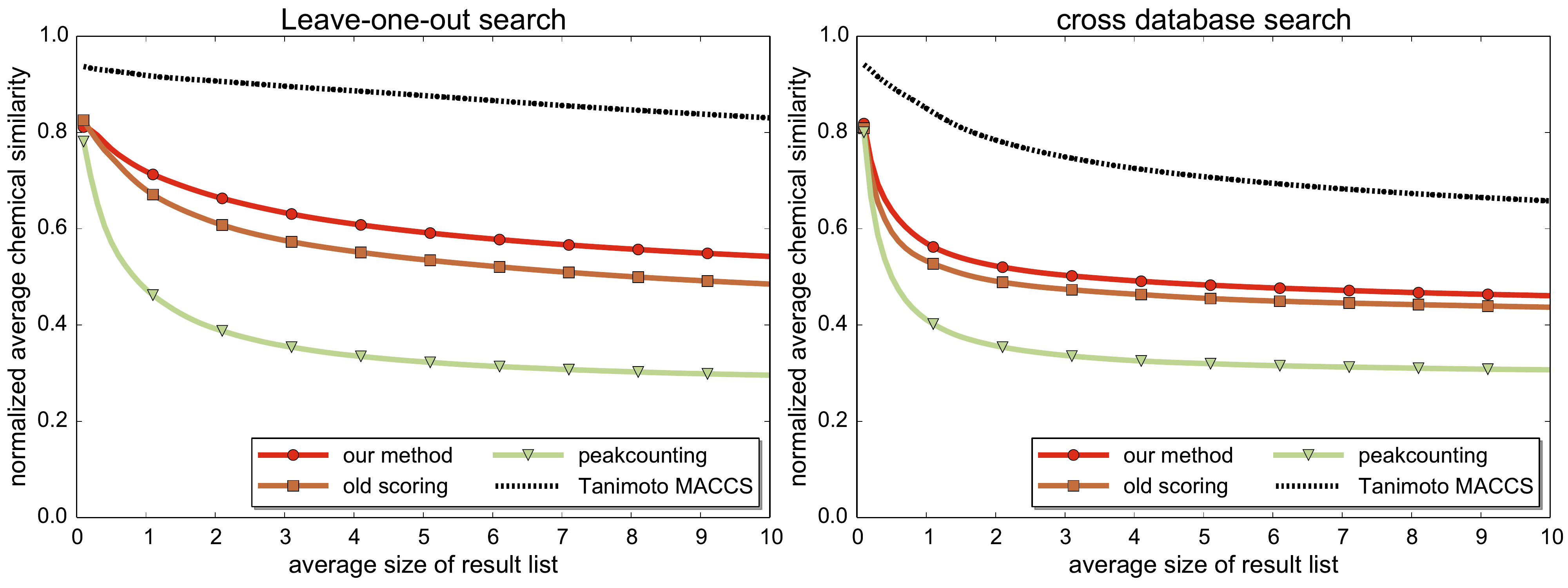}

\caption{Similarity search performance plots for chemical similarity. 
  Methods ``our method'' and ``old scoring'' compute FTs using ILP and
  compare trees via tree alignments~\cite{rasche12identifying}.  Method
  ``peak counting'' uses direct spectral comparison. Method ``MACCS'' uses fingerprints computed from the structure of the compound.  Left: Similarity search 
  with Leave-one-out strategy on both datasets.  Right: Similarity search across 
  the databases. Compounds from GNPS are searched in Agilent and vice-versa. }
\label{fig:structure-results}
\end{figure}

As our second evaluation of FT quality, we want to search a spectral library
with a query compound \emph{not} contained in the database; the goal of this
search is to find compounds that are structurally
similar~\cite{demuth04spectral, rasche12identifying}.  In a leave-one-out
evaluation, we use each compound as our query; for each query, we sort all
remaining entries of the database with respect to our similarity score, then
evaluate the average chemical similarity of the first $k$ entries. 
Instead of forcing each query compound to return the same number of entries,
we just enforce that in average each query returns $k$ entries. The cross database
evaluation is done analogously, but using GNPS compounds as query and searching in the Agilent database (and vice-versa).  

We measure chemical similarity using Tanimoto coefficients and PubChem
fingerprints.  See Fig.~\ref{fig:structure-results} for a comparison of the
old and new FTs, using tree alignments from \cite{rasche12identifying} to
compute similarity scores.  We also compare to direct spectral comparison via
peak counting, which gave us best results of all methods for direct spectral
comparison on these datasets, and Tanimoto scores computed by MACCS fingerprint.
Remark that for computing MACCS fingerprints, the structure of the compound have to be known, 
while for spectral and FT alignments only the spectrum is necessary. 
 We normalize score such that the optimal method reaches
similarity score $1$, and the random method reaches~$0$.

\section{Conclusion}

We have presented a maximum a~posteriori estimator for the problem of
computing fragmentation trees, that performs significantly better than
previous approaches for the problem.  Identification
performance can be significantly improved by adding isotope pattern
information~\cite{boecker09sirius, rasche11computing, duehrkop14molecular}
but this data is not available for the two datasets.  
The only alternative method for estimating a
molecular formula (solely) from tandem MS data is the commercial MOLGEN-MS/MS
software~\cite{scheubert13computational, meringer11msms}, which performs
roughly on par with SIRIUS$^2$ (DP version)~\cite{stravs13automatic}.

We used the new scoring in the CASMI (Critical Assessment of Small Molecule
Identification) challenge 2013 to determine the molecular formula of 12
unknown compounds.  Using the fragmentation tree analysis as presented here,
we correctly identified 8 molecular formulas, and placed an additional 3 in
the TOP2~\cite{duehrkop14molecular}.  In conjunction with isotope pattern
analysis~\cite{boecker09sirius} we identified 10 out of 12 molecular
formulas, and our method SIRIUS was selected ``best automated tool'' of the
molecular formula challenge~\cite{nishioka14winners}.  Furthermore, the new
scoring was used to compute fragmentation trees as part of a novel approach
for determining molecular fingerprints from tandem MS data which, in turn,
can be used to search molecular structure databases~\cite{shen14metabolite}.
Here, the improved FT structure resulted in significantly improved prediction
performance.

\paragraph{Acknowledgments.}

We thank Frank Kuhlmann and Agilent Technologies, Inc.\ (Santa Clara, USA)
for providing uncorrected peak lists of their spectral library.  We thank
Pieter Dorrestein, Nuno Bandeira (University of California) and the GNPS
community for making their data accessible.

\end{document}